\documentclass[%
superscriptaddress,
twocolumn,
 amsmath,amssymb,
 aps,
prb,
floatfix,
]{revtex4-2}

\usepackage{graphicx}
\usepackage{array}
\usepackage{dcolumn}
\usepackage{bm}
\usepackage{caption}
\usepackage{subcaption}
\usepackage{braket}
\usepackage[normalem]{ulem}
\usepackage{xcolor}
\usepackage{comment}

\begin{document}
\preprint{APS/123-QED}

\title{Interlayer Dzyaloshinskii-Moriya interactions induced via non-linear phononics in bilayer van der Waals materials}

\author{Ze-Xun Lin}
\email{zexun.lin@utexas.edu}
\affiliation{%
Department of Physics, The University of Texas at Austin, Austin, TX 78712, USA
}
\affiliation{Department of Physics, Northeastern University, Boston, MA 02115, USA }
\affiliation{Division of Physical Sciences, College of Letters and Science, University of California, Los Angeles, CA 90095, USA}

\author{Bowen Ma}
\affiliation{%
Department of Physics and HK Institute of Quantum Science \& Technology, 
The University of Hong Kong, Pokfulam Road, Hong Kong, China
}
\author{Wesley Roberts}
\affiliation{Department of Physics, Northeastern University, Boston, MA 02115, USA }

\author{Martin Rodriguez-Vega}%
 \affiliation{%
Department of Physics, The University of Texas at Austin, Austin, TX 78712, USA
}
 \affiliation{Department of Physics, Northeastern University, Boston, MA 02115, USA }

\author{Gregory A. Fiete}
\affiliation{Department of Physics, Northeastern University, Boston, MA 02115, USA }
\affiliation{Department of Physics, Massachusetts Institute of Technology, Cambridge, MA 02139, USA}



\date{\today}

\begin{abstract}
We theoretically study the impact of light-driven structural changes via nonlinear phononics on the magnetic order of untwisted bilayer van der Waals materials.  We consider an illustrative example of the AA-stacked bilayer honeycomb lattice and show that high-intensity light in resonance with selected phonons induces large amplitude phonon displacements that modify the magnetic Hamiltonian of the system. We performed a group theory analysis to identify the vibrational modes of the honeycomb bilayer and the nonlinear couplings among them in the strongly driven regime. We find that the  structural changes in the strongly driven regime lower the symmetry relative to the equilibrium lattice and produce changes in the magnetic interactions between the local moments. In particular, the lattice symmetry changes permit a non-zero interlayer Dzyaloshinskii-Moriya interaction that induces a magnetic state with canted local moments. Using a spin-wave analysis about the new magnetic configuration we study the corresponding changes in the magnon spectrum and identify a protocol for engineering topological band transitions using a combination of nonlinear phononics and an external magnetic field. Our work suggests a strategy to induce and control interlayer Dyzaloshinskii-Moriya interactions in a class of layered van der Waals materials, the effect of which is to modify the magnetic ground state, magnon dispersions, and related band geometric properties, including topological invariants. 
\end{abstract}

\maketitle


\section{\label{sec:Intro}Introduction}

The allowed magnetic exchange interactions on a particular lattice are constrained by symmetry.   
The Dzyaloshinskii-Moriya interaction (DMI),
\begin{equation}
H_{\rm DMI}=\sum_{i,j} \vec D_{ij} \cdot (\vec S_i \times \vec S_j),  
\label{eq:DMI_gen}
\end{equation}
where $\vec S_i$ is a spin on site $i$ in the lattice and $\vec D_{ij}$ is a lattice symmetry-determined vector depending on the positions of site $i$ and $j$, is permitted in the absence of a center of inversion\cite{DZYALOSHINSKY1958241,Moriya1960AnisotropicFerromagnetism}, either in bulk or at an interface \cite{RevModPhys.89.025006, RevModPhys.95.015003}. As seen from Eq.\eqref{eq:DMI_gen}, the DMI tends to favor spin configurations where local moments exhibit a perpendicular relative orientation, in contrast to the collinear orientations favored by Ising exchange interactions, and Heisenberg/XY exchange interactions in the absence of frustration. (Frustration can produce noncollinear ground state spin arrangements in spin Hamiltonians with a continuous spin rotational symmetry, such as occurs on the triangular lattice with nearest-neighbor antiferromagnetic Heisenberg interactions \cite{PhysRevLett.60.2531,PhysRevLett.69.2590}.)   Through competition with Heisenberg exchange interactions, the DMI can lead to interesting magnetic textures, such as skyrmions \cite{Fert_2017}, which could find applications in spintronics \cite{Dieny2020,10.1063/5.0072735,10.1063/1.5134474}. 

The majority of studies on magnetic systems with a DMI have focused on the intralayer DMI, but recent theoretical predictions \cite{PhysRevLett.122.257202}, followed by experimental realizations \cite{Han2019,FernandezPacheco2019,PhysRevLett.127.167202, PhysRevB.105.184405}, have shown the emergence of interlayer DMI (IL-DMI) in layered magnetic systems. Furthermore, experiments have demonstrated the possibility of controlling the IL-DMI in synthetic antiferromagnets via electric currents \cite{doi:10.1021/acs.nanolett.3c01709}, which could enable the manipulation of three-dimensional magnetic textures \cite{Kent2021}.  

In this work, we explore how to generate an IL-DMI with optical control, namely by irradiating quantum materials with lasers, which previously has been demonstrated to induce and control various ordered states \cite{Basov2017, RevModPhys.93.041002, Rodriguez-Vega2022QuantumEquilibrium,Bao2021Light-inducedMaterials, Chang2013ExperimentalInsulator, Merboldt2024ObservationGraphene}. In particular, the laser excitation of infrared lattice vibrations has allowed researchers to predict, induce and manipulate magnetic states \cite{Nova2017,Disa2020,Disa2023} (including magnetic order switching \cite{Stupakiewicz2021, Davies2024}), ferroelectric states \cite{PhysRevLett.118.197601,vonHoegen2018,doi:10.1126/science.aaw4911, Henstridge2022, li2019terahertz}, and enhance superconductivity in organic materials \cite{Mitrano2016PossibleTemperature, Cantaluppi2018,Budden2021,Rowe2023,PhysRevX.10.031028,PhysRevLett.127.197002}, cuprates \cite{Fausti2011Light-inducedCuprate,PhysRevB.90.100503, PhysRevB.89.184516,Hu2014,PhysRevX.12.031008} as well as in more traditional BCS systems \cite{Eckhardt2024TheorySuperconductivity}. An ultrafast symmetry switching utilizing intense terahertz light coupled with phonon has been demonstrated experimentally in Weyl semimetal as well \cite{sie2019ultrafast}. Additionally, theoretical proposals have shown that cavities, instead of lasers, could also lead to phonon-induced ordered electronic states \cite{PhysRevResearch.3.L032046, Dag2023CavityGraphene, Hubener2020EngineeringCavities, Curtis2019CavitySuperconductivity}.  The laser illumination strategy has the additional advantage of ultrafast (and reversible) control of the magnetic Hamiltonian.

Prior theoretical works have shown that light coupled directly to the electronic or spin degrees of freedom can induce and control the (intralayer) DMI in 2D magnetic materials described by the Kane-Mele-Hubbard model \cite{PhysRevB.100.060410}, and in multiferroics \cite{PhysRevLett.117.147202}. After analyzing crystallographic point groups in two-dimensional insulating magnets, a subsequent symmetry analysis showed that light-induced symmetry lowering universally results in a DMI \cite{PhysRevB.108.064420}. However, introducing a mechanism that allows one to control the intralayer and IL-DMI interaction via the lattice vibrations could bypass (or at least minimize) the heating effects associated with such direct laser-electron coupling. 
 
Generally, the symmetry criterion on the IL-DMI are given by Moriya's rules \cite{Moriya1960AnisotropicFerromagnetism}. As long as the rules do not forbid the appearance of an IL-DMI, it will be present. Here we list Moriya's rules for a non-zero DMI \cite{Moriya1960AnisotropicFerromagnetism}, following Moriya's notation. Considering the coupling between two ions in the crystal, the positions of these two ions are denoted as position A and B, and the midpoint of AB is denoted as C. Moriya showed the following symmetry rules apply:
\begin{enumerate}
    \item When a center of inversion is located at C, then $\bm D=0$.
    \item When a mirror plane perpendicular to AB passes through C, then $\bm D \parallel$ mirror plane or $\bm D \perp$ AB.
    \item When there is a mirror plane including A and B, then $\bm D$ is $\perp$ to the mirror plane. 
    \item When a two-fold rotation axis  perpendicular to AB passes through C, then $\bm D$ is $\perp$ to the two-fold axis.
    \item When there is an n-fold axis ($n\geq 2$) along AB, then $\bm D$ is $\parallel$ to AB.  
\end{enumerate}

Therefore, the question remains of how to show how one can use a nonlinear phononics protocol to break symmetry such that Moriya's rules allow the IL-DMI in the new, non-equilibrium lattice configuration. For demonstration purposes, here we consider an insulating AA-stacked bilayer honeycomb lattice with localized classical moments with collinear order from Heisenberg interactions. In equilibrium, symmetry considerations forbid an IL-DMI between nearest neighbors. We show that applying an intense enough laser in resonance with specific infrared active phonons lowers the symmetry of the non-equilibrium structure via non-linear coupling with Raman active phonons. The out-of-equilibrium lattice structure permits a non-zero IL-DMI between nearest neighbors, which, in turn, leads to a canted magnetic state. We also analyze the corresponding changes to the magnon spectrum as well as the magnon band topology.    

The remainder of the paper is organized as follows. In Sec.\ref{sec:model}, we introduce a bilayer honeycomb lattice model and perform a symmetry (group theory) analysis to identify the normal modes of lattice vibrations (phonons) and their non-linear interactions. We then select the phonons that break the symmetry which forbids the existence of IL-DMI in equilibrium to allow the IL-DMI in the out-of-equilibrium lattice configuration. In Sec.~\ref{sec:Hamiltonian}, we introduce the equilibrium magnetic Hamiltonian and the form of the phonon-induced IL-DMI. We perform a spin-wave (magnon) analysis showing the effects of the IL-DMI. We suggest possible material candidates to observe the effects discussed in our work. Finally, in Sec.~\ref{sec:conclusion}, we present the main conclusions of our work.  Some technical details and figures are relegated to the appendices.

\section{IL-DMI from nonlinear phononics}
\label{sec:nonlinear}

\subsection{Brief review of nonlinear phononics}
We begin our discussion by studying phonons, which are quantized modes of lattice vibrations. In a system with inversion symmetry they can be categorized into two types. One type is infrared active (IR) modes which are directly related to the electric dipole moment and thus can be directly excited by an electric field with the correct frequency. The other type is Raman modes, which are related to the polarizability of the phonon mode. In a centrosymmetric crystal, only infrared active modes can be directly controlled by an electric field, while Raman modes require a second order photon process to excite \cite{Dresselhaus2010GroupMatter}. However, Raman active modes can be controlled indirectly via non-linear couplings to infrared active modes - an approach called nonlinear phononics \cite{Henstridge2022,Forst_2011}. In particular, the nonlinear couplings between the two modes can be used to shift the equilibrium position of the Raman modes which are typically lower in frequency by looking at the average effect of the ``fast" IR active modes. This approach allows one to modify lattice symmetries and enable new forms of magnetic exchange terms in a local moment Hamiltonian on the lattice. 

In our work we will assume a frozen phonon picture as well as a Born-Oppenheimer approximation \cite{Born1927ZurMolekeln} for electrons responding to the transient lattice changes from light. This is well justified based on the light mass of the electrons relative to the lattice ions as well as the characteristic frequencies of phonons compared to electron energies (phonon frequencies are typically two orders of magnitude or more smaller than the eV energy scale).  We also treat the interaction between light and infrared active modes, as well as the interaction between phonon modes themselves, as classical. 

One can understand the qualitative effects from nonlinear phononics in a simple heuristic model:  When one takes into account up to cubic order terms in the lattice displacements the effective potential of a single interacting IR and Raman phonon reads \cite{Subedi2014TheorySolids,Radaelli2018BreakingCoupling},
\begin{eqnarray}
     V_{eff}&=&\frac{1}{2}\Omega_R^2Q_R^2+\frac{1}{2}\Omega_{IR}^2Q_{IR}^2-\frac{1}{4}gQ^2_{IR}Q_R\nonumber\\
     & &-F\Phi(t)\sin(\Omega t)\, Q_{IR},
     \label{eq:V_eff}
 \end{eqnarray}
 where $Q_{IR}$ and $Q_{R}$ are infrared and Raman lattice vibrational mode amplitudes, and $\Omega_R$ and $\Omega_{IR}$ are their corresponding frequencies. We assume that incident laser light on the material has a Gaussian envelope intensity $\Phi(t)\propto e^{-t^2/2\sigma^2}$ where $\sigma$ is the characteristic value for the pulse length. The strength of an incoming laser field is determined by the parameter $F=\mathcal{Z}^*\cdot E$, where $\mathcal{Z}^*$ is the so-called Born effective charge~\cite{Radaelli2018BreakingCoupling} and $E$ the peak electric field amplitude of the laser pulse. The parameter $g$ is the coupling constant between Raman and infrared modes, appearing in the third term of Eq.\eqref{eq:V_eff}.

 In the impulsive limit $\Omega\sigma \ll 1$, solving the equation of motion for the infrared mode, one has~\cite{Subedi2014TheorySolids},
\begin{equation}
     Q_{IR}=-\sqrt{2\pi}F\Omega_{IR}\sigma^3\cos\Omega_{IR}t,
\end{equation}  
which shows the IR active mode oscillates with frequency $\Omega_{IR}$. In the same limit the equilibrium position of the Raman mode is displaced as $Q_R\rightarrow Q_R+\Delta Q_R$ \cite{Subedi2014TheorySolids}, where
\begin{equation}
     \Delta Q_{R}=\frac{\pi}{2}\left(\frac{\Omega_{IR}}{\Omega_R}\right)^2gF^2\sigma^6.
\end{equation}  
Therefore, the equilibrium position of the Raman mode can be shifted by an amount depending on the square of the ratio of IR to Raman frequencies, $\left(\frac{\Omega_{IR}}{\Omega_R}\right)^2$, and the square of the force, $F^2$, acting on the IR mode. Broader pulses lead to a larger Raman equilibrium shift, $\Delta Q_{R}$.  These principles were recently shown to result in coherent excitation of phonon modes in La$_{0.7}$Sr$_{0.3}$MnO$_3$\cite{Forst2011NonlinearControl}, changing ferroelectric polarization at the surface of LiNbO$_3$\cite{Henstridge2022NonlocalPhononics},  forming ferrimagnetic order in CoF$_2$\cite{Disa2020PolarizingField} and a dynamical control of interlayer magnetic exchange coupling in bilayer CrI$_3$ \cite{Rodriguez-Vega2020Phonon-mediatedI3} and bilayer MnBi$_2$Te$_4$~\cite{Rodriguez-Vega2022Light-DrivenAntiferromagnets},  as well as a topological band transition in the latter\cite{Rodriguez-Vega2022Light-DrivenAntiferromagnets}. It has also been shown to be applicable in enhancing superconductivity in organic materials \cite{Mitrano2016PossibleTemperature, Cantaluppi2018,Budden2021,Rowe2023,PhysRevX.10.031028,PhysRevLett.127.197002}, cuprates \cite{Fausti2011Light-inducedCuprate,PhysRevB.90.100503, PhysRevB.89.184516,Hu2014,PhysRevX.12.031008} as well as in more traditional BCS systems \cite{Eckhardt2024TheorySuperconductivity}.

\subsection{Nonlinear phononics in a bilayer honeycomb lattice}
\label{sec:model}
Group theory constrains the real space displacement of the phonon modes and their effective potential, which generally contains anharmonic terms. The group theory perspective provides an intuitive understanding of how the magnetism is related to lattice symmetry and thus how it can be directly controlled by the nonlinear phononics mechanism. A transient change in the lattice structure from the nonlinear phononics mechanism changes the inter-atomic hopping parameters for electrons, which in turn produces a modification in the magnetic exchange interaction.

This article focuses on a bilayer honeycomb lattice system to illustrate the physics in a simple yet relevant setting. Specific to this system, a symmetry change of the lattice induced by non-linear phononics produces an IL-DMI which is absence in the equilibrium lattice structure.  This IL-DMI produces a non-collinear ground state with canting and results in topological magnon bands for a certain range of parameters.

The honeycomb lattice model we study in this paper is motivated by a broad class of materials with a layered honeycomb lattice structure. Common examples are the transition-metal dichalcogenides VX$_2$(X=S, Se, and Te), RuI$_3$\cite{Banerjee2018Excitations-RuCl3}, RuCl$_3$, and MoS$_2$. For simplicity, we consider a honeycomb bilayer as a minimal model. Before we examine the effect that Raman modes can have on the DM interaction, we will study which vibrational modes are possible in a honeycomb bilayer. There are two most common stacking orders for a bilayer honeycomb lattice, referred to as AA and AB stacking (Bernard stacking). The point group for AA stacking is $D_{6h}$ while the point group for AB stacking is $D_{3d}$. Both symmetries enforce the nearest neighbor interlayer DM interaction to be zero in equilibrium~\cite{Moriya1960AnisotropicFerromagnetism}. In bilayer CrI$_3$, there is a third type of stacking order, called AB', with point group $C_{2h}$, which is a high-temperature bulk stacking. This structure allows a nonzero interlayer DMI even in equilibrium, as we discusss in Appendix~\ref{third_stacking}.

Both AA and AB stacked bilayer honeycomb lattices have twelve phonon modes since there are four atoms in the unit cell and three spatial directions in which atoms can move. In the case of AB stacking, the inversion center of the bilayer is the same as the nearest neighbor interlayer bond center, and thus all Raman modes are inversion symmetric at the bond center, forbidding the emergence of the nearest-neighbor IL-DMI. (See Appendix \ref{append:GroupTheory} for details.) 
Although, thermodynamically, for most materials, AB stacking is energetically more favorable \cite{Nguyen2020Layer-controlledFormation}, only in the case of AA stacking can one generate IL-DMI between the layers through non-linear phononics.  

In AA stacking, the system has $D_{6h} = D_6 \otimes i$ point group symmetry where $D_6$ has the symmetry operations $\{E, 2C_6, 2C_3, C_2, 3C'_2, 3C''_2\}$ and $i$ represents inversion symmetry. The irreducible representations are $A_{1g(u)}$, $A_{2g(u)}$, $B_{1g(u)}$, $B_{2g(u)}$, $E_{1g(u)}$, $E_{2g(u)}$.  The $g(u)$ subscripts represent modes even (Raman) or odd (IR) under inversion symmetry. We now compute the irreducible representations of the vibrational modes in AA stacking, given by $\chi_{vibration}=\Gamma^{equivalence}\otimes\Gamma_{vec}=A_{1g}\oplus A_{2u}\oplus B_{2g}\oplus B_{1u}\oplus E_{2g}\oplus E_{1u}\oplus E_{1g}\oplus E_{2u}$,
 where $\Gamma^{equivalence}=\Gamma^{\text{atom site}}$ keeps track of the number of atoms that are mapped onto their same positions under point group operations, and $\Gamma_{vec}$ is the vector representation whose basis functions are $x,y,z$ \cite{Dresselhaus2010GroupMatter}. 
 


\begin{figure}[ht]
\includegraphics[width=40mm]{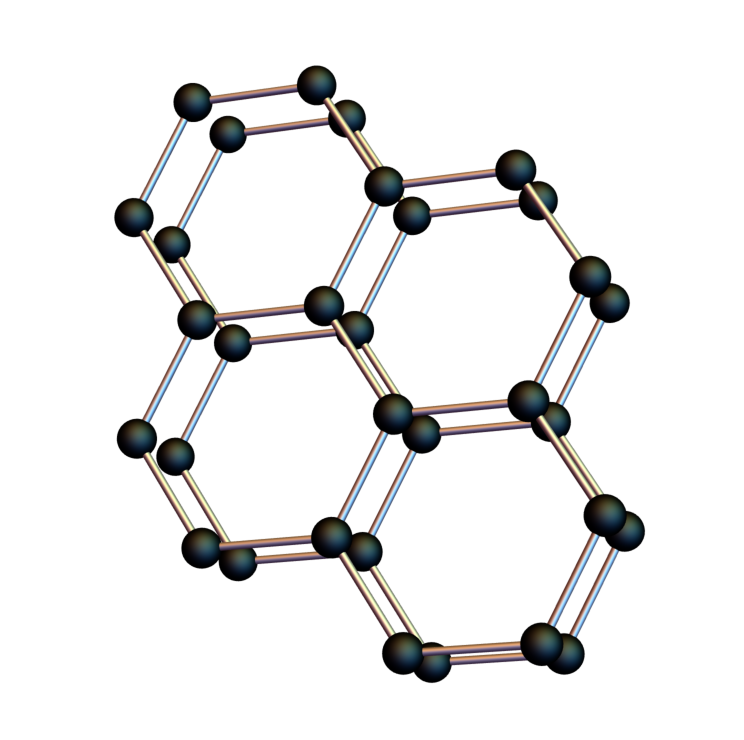}
\caption{ AA stacking of a bilayer honeycomb lattice.  Each atom sits either directly above or below an atom in the adjacent layer.}
\label{fig:abstacking}
\end{figure}
 
Three of these modes are acoustic: the $A_{2u}$ mode and both $E_{1u}$ modes, which correspond to center-of-mass (COM) motion in the $z,x,y$ directions. Besides these COM modes, there are modes that can break crucial symmetries enabling the existence of interlayer DM interactions. Among them, singly degenerate modes $A_{1g}$ and $B_{2g}$ exhibit out-of-plane motions while doubly degenerate $E_{1g}$ and $E_{2g}$ modes have in-plane motion only. In this article, we will focus on the $E_{2g}$ modes since the $E_{1g}$ mode is a shearing motion between layers, whose effect has been explored in an earlier work \cite{Rodriguez-Vega2022Light-DrivenAntiferromagnets}.

\begin{figure}[ht!]
\includegraphics[width=40mm]{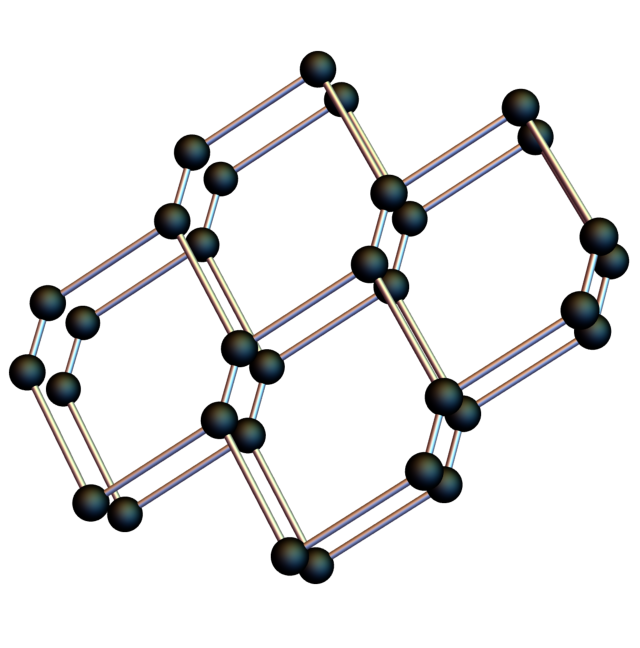}
\caption{\label{fig:E2g} $E_{2g}$ oscillation in the $y$-direction for an AA stacked honeycomb bilayer, effectively changing the bond length in $y$-direction}
\end{figure}

To determine the corresponding real space displacements (up to a unitary change of basis), we use the projection operator \cite{Dresselhaus2010GroupMatter}
 \begin{equation}
    \hat P^{(\Gamma_n)}_{kl} = \frac{l_n}{h} \sum_{C_\alpha} \left( D_{kl}^{(\Gamma_n)}(C_\alpha) \right)^* \hat P(C_\alpha), 
 \end{equation}
where $D_{kl}^{(\Gamma_n)}(C_\alpha)$ is the irreducible matrix
representation of the group element $C_\alpha$, $h$ is the order of the group, $l_n$ is the dimension of the irreducible representation, and $\hat P(C_\alpha)$ is the representation of $C_\alpha$ constructed by the permutation matrix and the O$(3)$ symmetry operations. The computed real space displacements of the $E_{2g,y}$ mode are shown in Fig.\ref{fig:E2g}, which change the bond length in the $y$-direction. The degenerate partner $E_{2g,x}$ changes the bond length in the $x$-direction.

The $E_{2g,y}$ modes break the 3-fold rotation axis which allows an inter-layer DM interaction, with the symmetry requirement $\vec D \perp yz$-plane, where $\vec D$ appears in Eq.\eqref{eq:DMI_gen}. The character table (from group theory) shows the $E_{2g}$ mode has quadratic generating functions: $x^2-y^2$ and $xy$, which identifies it as a Raman mode. The $E_{1u}$ mode has linear generating function $(x,y)$ identifying it is an IR mode.

By a group theory analysis \cite{Dresselhaus2010GroupMatter}, symmetry-allowed anharmonic coupling between the $E_{1u}$ IR mode $Q_{IR}$ and the $E_{2g}$ Raman mode $Q_{R}$ can be written as,
\begin{equation}
   V_{anharmonic}=({Q_{IR}^y}^2-{Q_{IR}^x}^2)Q_{R}^a+2Q^x_{IR}Q^y_{IR}Q_{R}^b.
\end{equation}
where the ``a" and ``b" refer to orthogonal vibration modes, which could be taken as a linear combination of $x$ and $y$-directional vibrations. The average positions of Raman modes when coupled to the IR modes with an external electric field are shifted as \cite{Radaelli2018BreakingCoupling, PhysRevLett.118.054101},
 \begin{eqnarray}
Q_{R}^a&\propto&\left(\frac{\Omega_{IR}}{\Omega_{R}}\right)^2(E^2_y-E^2_x),\label{eq:mode_shift_a}\\
Q_{R}^b&\propto&\left(\frac{\Omega_{IR}}{\Omega_{R}}\right)^2(2E_xE_y\cos\Delta\phi),
\label{eq:mode_shift_b}
\end{eqnarray}
where $E_x$ and $E_y$ are the electric field components in the $x$ and $y$-directions which couple to the electric dipole of the IR modes, and $\Delta\phi$ is the phase difference between the $x$ and $y$-components. Detailed calculations are given in Radaelli \cite{Radaelli2018BreakingCoupling}.


From Eq.\eqref{eq:mode_shift_a} and Eq.\eqref{eq:mode_shift_b}, it is evident that one can excite Raman modes in a specific direction through a choice of the electric field direction. In the Sec.~\ref{E2G}, we will see how these modes influence the intra-layer DM interaction and spin ground state. Note that in the usual experimental situation, most bilayer structures energetically favor AB stacking, but under pressure and with topological defects \cite{Mcclarty2021TopologicalReview}, one can still achieve AA stacking.


 \subsection{Nonzero DMI from a driven $E_{2g}$ mode}\label{E2G}
The key observation is that if one excites the $E_{2g}$ Raman mode in the $y$-direction, the point group of the AA stacked bilayer effectively changes from $D_{6h}$ to  $D_{2h}$. Taking the time average of the Hamiltonian over a period of the Raman oscillation, the nonlinear phononics coupling effectively changes the lattice configuration, and allows a non-zero DM vector, $\vec D$, for nearest neighbor inter-layer exchange interactions. This is one of the central results of this work. The DMI originates from the breaking of inversion symmetry along the bond connecting nearest neighbors in the upper and lower layer of the AA stacked honeycomb lattices.

Focusing on the interlayer DM interaction in the static case, it would be zero due to Moriya's rules: it has both a 3-fold rotational axis along the bond and a mirror plane including the bond. Breaking of $C_3$ symmetry would lower the symmetry to only mirror symmetry, allowing a DMI perpendicular to the mirror plane. To linear order in the Raman mode displacement,
\begin{equation}
    |\vec D|\propto \lambda Q_R d,
\end{equation}
where $d$ is the distance separating the two layers, and $\lambda$ is the strength of the spin-orbit coupling. The direction of $\vec D$ is perpendicular to the $yz$-plane. An exact relation between the interlayer DM and phonon modes would need further detailed first principle calculations \cite{yang2023first}\cite{stavric2023delving} for a specific material.  We are interested primarily in general considerations here.

In the following section, we will show that this light-induced IL-DMI can change the magnetic order of the ground state and can result in a gap opening for magnetic excitations, {\em i.e.} the magnons, above the ground state, leading to a topological magnon band transition.

\section{Model Spin Hamiltonian}
\label{sec:Hamiltonian}
The above discussion on driven phonons  depends only on lattice symmetry rather than an explicit spin Hamiltonian. To explore the consequences of the generation of an IL-DMI we choose a spin Hamiltonian with an easy-axis exchange interaction, where the intra-layer exchange interaction is ferromagnetic and the inter-layer exchange is chosen to be either ferromagnetic or anti-ferromagnetic to represent different possibilities for material realizations.

\begin{figure}
    \centering
\includegraphics[width=0.4\textwidth]{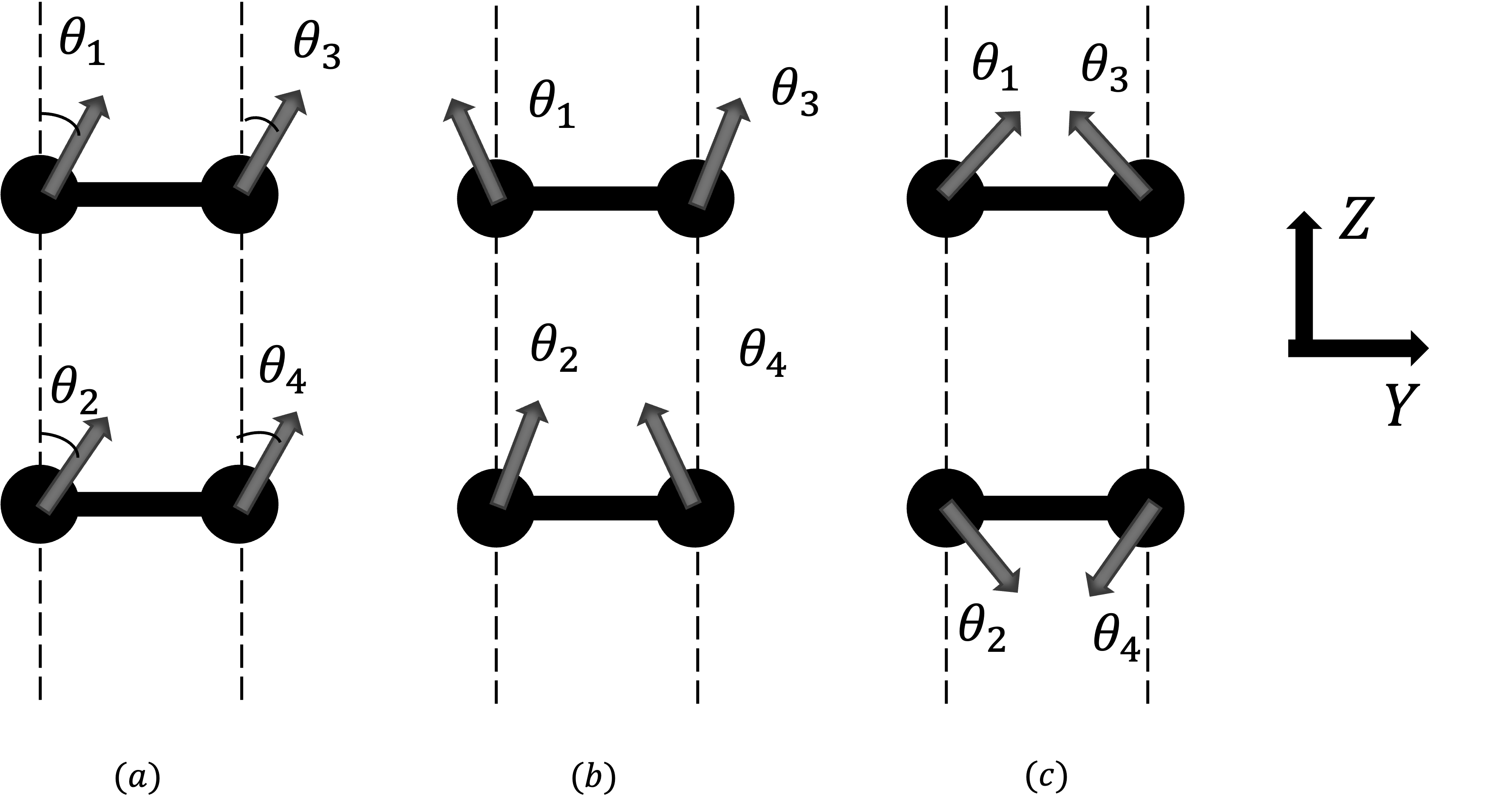}
    \caption{The ground state magnetic configuration showing the spin orientation of the two basis atoms in the unit cell for the upper and lower layers of the honeycomb bilayer. The IL-DMI vector $\vec D$ is perpendicular to the $ yz$ plane. Here, (a) shows the angles' sign convention. (b) shows the IL-DMI produces a coplanar canting of the spins in the $ yz$ plane when one has FM interlayer exchange interaction. (c) shows the IL-DMI produces a coplanar canting of the spins in the $yz$-plane, when one has AFM interlayer exchange interaction. }
        \label{fig:canting}
\end{figure}

In the presence of the symmetry lowering from the $E_{2g}$ shear mode described in the previous section, an IL-DMI and bond-dependent nearest-neighbor exchange interactions are induced. The effective light-driven spin Hamiltonian can be written as,
\begin{eqnarray}
H&=&\sum_{\langle i,j\rangle,\eta}\left[ J_1\vec S_{i,\eta}\cdot \vec S_{j,\eta}-\Delta (S_{i,\eta}^z )^2\right]\nonumber\\
&+&\sum_{\langle \langle i,j\rangle\rangle ,\eta}\left[D_z\hat{z}\cdot \left(\vec S_{i,\eta}\times \vec S_{j,\eta}\right)\right]\nonumber\\
&+&\sum_{i} \left[ J_2 \vec S_{i,b}\cdot \vec S_{i,t}+\vec D_i \cdot( \vec S_{i,b}\times \vec S_{i,t})\right],
\label{eq:Hamiltonian}
\end{eqnarray}
where $\eta=t,b$ is the top or bottom layer index, $i,j$ are site indices, and $J_1<0$ corresponds to ferromagnetic interactions within the plane. Here $J_2<0$ corresponds to the ferromagnetic coupling between layers, and $J_2>0$ antiferromagnetic coupling. We also include a second nearest-neighbor intralayer DMI $D_z \hat{z}$ that is generally allowed by the symmetry, even in the equilibrium configuration.  As indicated in Fig.\ref{fig:canting}, $\vec D_i$ in the last term of Eq.\eqref{eq:Hamiltonian} is perpendicular to the $yz$-plane. We choose a convention ($\Delta <0$) which favors all spins pointing in the $z$-direction for simplicity. In principle, all the exchange parameters can change when driving the system, but we focus on the IL-DMI term, $\vec D_i \cdot( \vec S_{i,b}\times \vec S_{i,t})$, which becomes non-zero upon lowering the lattice symmetry. 

To obtain a better sense of the scale of the exchange interactions for some representative materials, note that $|J_2|/|J_1|$ is about 0.004 and 0.06 in CrI$_3$ and CrBr$_3$, respectively \cite{Owerre2017DiracMagnets}.  The lowering of the symmetry via nonlinear phonics produces the new IL-DMI term which changes the symmetry of the magnetic ground state.  The interlayer DM must be perpendicular to the $y$-axis and parallel to the $xy$-plane, so IL-DMI takes the following form: $\vec D_i=(D,0,0)$. The sign of $D$ depends on material details and is related to the direction of the effective electric field contributing to the spin-orbit coupling \cite{Moriya1960AnisotropicFerromagnetism}.  

Since the DMI, which originates from SOC, is usually one order of magnitude smaller than $J_1$, we assume the DMI only contributes small canting to the original collinear ordering rather than producing spiral orders or other spin textures that enlarge the unit cell. The lowest-energy magnetic configuration can be calculated classically under the ansatz that the magnetic order does not enlarge the unit cell by minimizing the classical energy for different configurations of the magnetic moments. The lowest-energy configuration, when no external magnetic field is added, is illustrated in Fig.~\ref{fig:canting}. The canting structure is coplanar when there is no external field. When the interlayer coupling is ferromagnetic ({\em i.e.}, $J_2<0$), $\theta_1=\theta_4=-\theta_3=-\theta_2$. For the AFM interlayer coupling case ({\em i.e.}, $J_2>0$), $\theta_1=-\theta_3=\pi-\theta_2=\theta_4-\pi$. Applying an external magnetic field for directions not in the plane of the canting makes the spin stucture non-coplanar.

\subsection{Magnon spectrum in the driven lattice}
\label{sec:magnon_spectrum}
To compute the magnon spectrum we perform a Holstein-Primakoff transformation \cite{Holstein1940FieldFerromagnet} in the non-interacting magnon limit which is the leading order in a $1/S$ expansion for fluctuations around the lowest-energy magnetic configuration, where $S$ is the magnitude of the spin.  For the case of a non-collinear ground state, before expanding in the magnon basis, one needs to perform a local spin-rotation transformation so that the local $z$-axis points along the direction of the spin in the classical ground state. To express the spin orientations resulting from the classical energy minimization, we use the standard cylindrical coordinate system,
\begin{equation}
    \vec S_{i}= S(\sin \theta_i \cos(\pi/2+\phi_i), \sin\theta_i \sin(\pi/2+\phi_i), \cos\theta_i).
\end{equation}

In this way, we rewrite the spin Hamiltonian to a local basis such that we could perform the Holstein-Primakoff transformation on a local basis.
\begin{equation}
    \begin{pmatrix}
        S_x\\
        S_y\\
        S_z
    \end{pmatrix}=R_z(\frac{\pi}{2}+\phi_i)R_x(-\theta_i) \begin{pmatrix}
        S_{x_L}(i)\\
        S_{y_L}(i)\\
        S_{z_L}(i)
    \end{pmatrix},
\end{equation}
where $R_x$ and $R_z$  are the standard rotational matrices in three-dimensional space.  The values of $\theta$ and $\phi$ are determined through classical energy minimization.

The spin component in the local coordinates are related to Holstein-Primakoff bosons as,
 \begin{eqnarray}
S_L^+&=&\hbar\sqrt{2S}\sqrt{1-\frac{b^\dagger b}{2S}}b,\\\nonumber
&\approx&\hbar\sqrt{2S}b,\\
S_L^-&=&\hbar\sqrt{2S}b^\dagger\sqrt{1-\frac{b^\dagger b}{2S}},\\\nonumber
&\approx&\hbar\sqrt{2S}b^\dagger,\\
     S_L^z&=&\hbar(S-b^\dagger b),
 \end{eqnarray}
where $b$ are the Holstein-Primakoff bosons, and the sublattice bases should be thought of as carried implicitly. The canting induced from the IL-DMI breaks the spin $U(1)$ symmetry of the collinear state down to $\mathbb Z_2$, leading to the pairing of bosons. Therefore, one needs to introduce the Bogoliubov de Gennes (BdG) basis~\cite{Bogoljubov1958OnSuperconductivity},
\scalebox{0.8}{$\Psi^\dagger_{\vec{k}}=(b^\dagger_{At,\vec{k}}, b^\dagger_{Bt,\vec{k}}, b^\dagger_{Ab,\vec{k}}, b^\dagger_{Bb,\vec{k}}, b_{At,-\vec{k}}, b_{Bt,-\vec{k}},b_{Ab,-\vec{k}},b_{Bb,-\vec{k}})$}. After a Fourier transformation to reciprocal space, the magnon Hamiltonian can be written in BdG form as,
\begin{equation}
    H_{\rm magnon}=\int_{BZ}d\vec{k}\Psi^\dagger_{\vec{k}} H_{BdG}(\vec{k})\Psi^{}_{\vec{k}}.
\end{equation}
In the absence of an external magnetic field, the groundstate satisfies $\theta_3=\theta_4=-\theta_1=-\theta_2=-\theta$ and $\phi_{1(2,3,4)}=0$, the explicit form of the BdG Hamiltonian is 
\begin{footnotesize}
\begin{equation}
\begin{aligned}
    H_{BdG}&(\vec k)=-\frac{1}{4}\left[ \Delta+(6J_1-2J_2+3\Delta)\cos(2\theta)  \right]+\frac{D}{2}\sin(\theta)\tau_z\xi_y\\
    &+J_2\cos(\theta)^2\tau_x\xi_x+\frac{J_2}{2}\sin(\theta)^2\xi_x-\Delta\sin(\theta)^2\tau_x\\
    &+f_1(\vec k)J_1\cos(\theta)^2\sigma_x-f_2(\vec k)J_1\cos(\theta)^2\sigma_y\\
    &+f_1(\vec k)J_1\sin(\theta)^2\tau_x\sigma_x-f_2(\vec k)J_1\sin(\theta)^2\tau_x\sigma_y\\
 &+D_z \cos(\theta)f_3(\vec k)\tau_z\xi_z\sigma_z,
\end{aligned}
\end{equation}
\end{footnotesize}
where $\tau$, $\xi$ and $\sigma$ are matrices in particle-hole, layer, and sub-lattice subspaces, respectively. The lattice basis vectors are chosen as $\vec R_1=-\frac{a}{2}\vec i+\frac{\sqrt{3}a}{2}\vec j$, $\vec R_2=\frac{a}{2}\vec i+\frac{\sqrt{3}a}{2}\vec j$, and sublattice basis is $\vec R_{AB}=\frac{\sqrt{3}a}{3}\vec j$
, also $f_1(\vec k)=Re[f(\vec k)]
$, and $f_2(\vec k)=Im[f(\vec k)]
$, $f_1+if_2=f(\vec k)=\frac{1}{2}\sum_i e^{i\vec k\cdot \vec a_i}$, $\vec a_i$ are the three NN bonds $\vec a_1=-\frac{\sqrt{3}a}{3}\vec j$, $\vec a_2=\frac{a}{2}\vec i+\frac{\sqrt{3}a}{6}\vec j$, $\vec a_3=-\frac{a}{2}\vec i+\frac{\sqrt{3}a}{6}\vec j$,  $f_3(\vec k)= Im[\sum_i e^{i\vec k \cdot \vec b_i}]=\left(2 \cos\left(\frac{\sqrt{3} k_y}{2}\right) \sin\left(\frac{k_x}{2}\right) - \sin(k_x)
\right)$, $\vec b_i$  are the three NNN bonds, $\vec b_1=\frac{a}{2}\vec i-\frac{\sqrt{3}a}{2}\vec j$, $\vec b_2=\frac{a}{2}\vec i+\frac{\sqrt{3}a}{2}\vec j$, $\vec b_3=-a\vec i$.

In determining the magnon band structure and eigenstates, one must take into account the bosonic statistics of the magnons,
\begin{equation}
[\Psi^{}_{\vec{k}},\Psi^\dagger_{\vec{k}}]={\rm diag}(1,1,1,1,-1,-1,-1,-1)\equiv \Sigma_z.
\end{equation} To diagonalize the magnon BdG Hamiltonian without changing the bosonic commutation relation, we use a para-unitary matrix $T_{\vec{k}}$~\cite{Kondo2020Non-HermiticitySystems}, such that the BdG Hamiltonian can be diagonalized by the matrix $T_{\vec{k}}$ as 
\begin{small}
\begin{equation}
	      T_{\vec{k}}^\dagger H_{BdG}(\vec k)T_{\vec{k}}=\begin{pmatrix}
	          E_{1,\vec{k}} & & & & &\\
            & \ddots & & & &\\
           & & E_{4,\vec{k}} & & &\\
           & & &E_{1,-\vec{k}} & &\\
           & & & & \ddots &\\
           & & & & & E_{4,-\vec{k}}\\
	      \end{pmatrix},
\end{equation}
\end{small}
with
\begin{equation}
	    T^{}_{\vec{k}}\Sigma_z T_{\vec{k}}^\dagger=\Sigma_z,
	  \end{equation}
to preserve the bosonic commutator.

Because of the particle-hole symmetry of the BdG basis, $E_{n,\vec{k}}\ (E_{n,-\vec{k}})$ is the eigenenergy of the $n^{th}$ particle (hole) band. Since $T_{\vec{k}}^\dagger=\Sigma_z T^{-1}_{\vec{k}}\Sigma_z$, we have
\begin{small}
\begin{equation}
	      T_{\vec{k}}^{-1}\Sigma_z H_{BdG}(\vec k)T_{\vec{k}}=\scalebox{0.85}{$\begin{pmatrix}
	          E_{1,\vec{k}} & & & & &\\
            & \ddots & & & &\\
           & & E_{4,\vec{k}} & & &\\
           & & &-E_{1,-\vec{k}} & &\\
           & & & & \ddots &\\
           & & & & & -E_{4,-\vec{k}}\\
	      \end{pmatrix}$}.
\end{equation}
\end{small}
With this transformation, the eigenproblem of solving the BdG Hamiltonian can be reduced to solving the eigenproblem of the non-Hermitian Hamiltonian $\Sigma_z H_{BdG}$. 

\begin{figure}[htp]
    \centering
\includegraphics[width=0.48\textwidth]{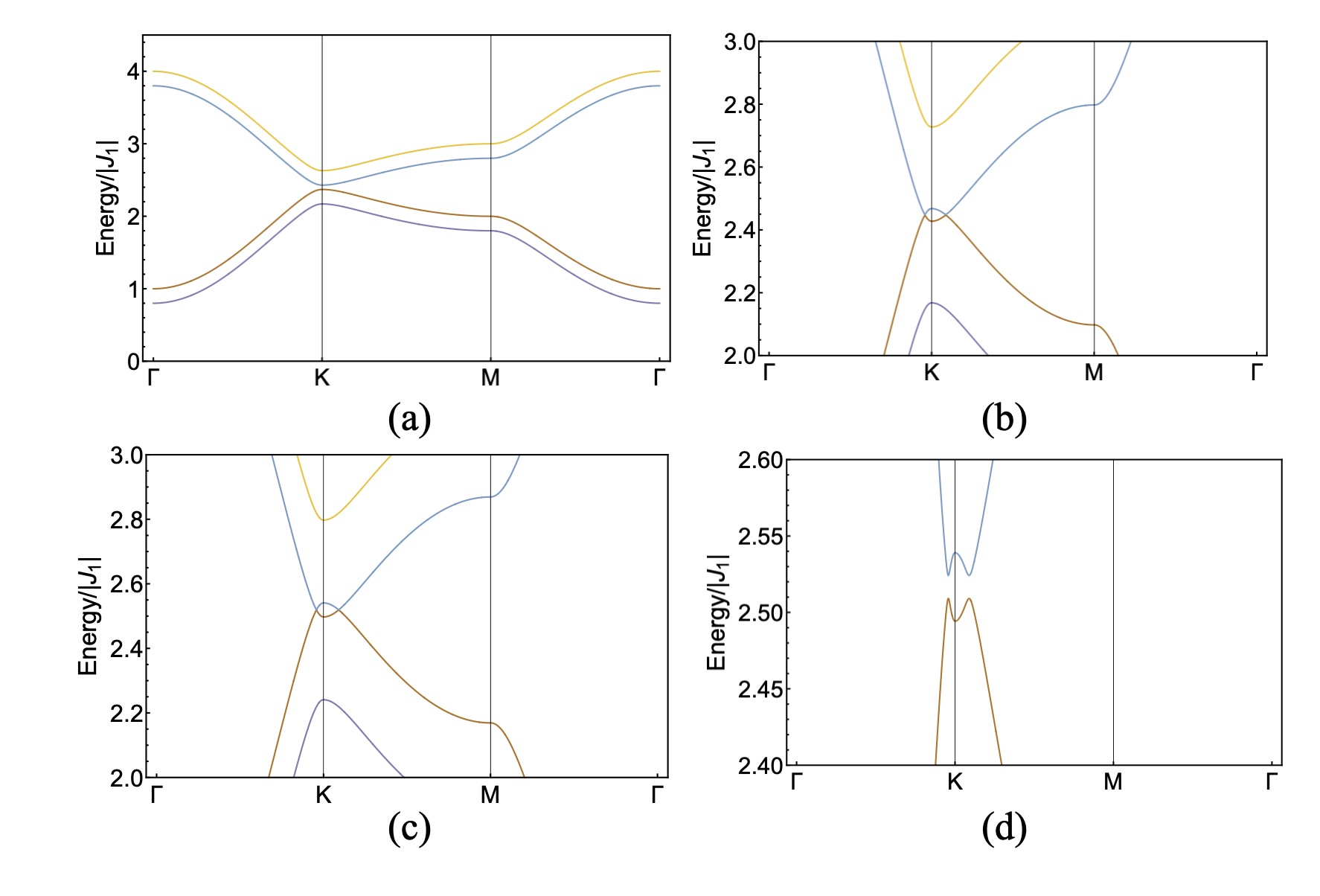}
    \caption{Magnon spectrum with FM interlayer coupling, other model parameters are: intralayer exchange couplings $J_1=-1$, easy axis anisotropy $\Delta=-0.8$, intralayer DMI strength $D_z=0.05$. (a) Interlayer FM $J_2=-0.2$  $\vec D_i=-0.2$, $\vec B=(0,0,0)$. (b) Interlayer FM $J_2=-0.3$, $\vec D_i=-0.2$, $\vec B=(0, 0, 0)$. (c) Interlayer FM $J_2=-0.3$, $\vec D_i = 0 \hat{x}$, $\vec B=(0.2, 0.2, 0.2)$. (d) Interlayer FM $J_2=-0.3$, $\vec D_i = -0.2 \hat{x}$, $\vec B=(0.2, 0.2, 0.2)$ show a gap opening between the two middle bands when we have a combination of both the IL-DMI and an external magnetic field. }
        \label{fig:FM_band}
\end{figure}

In Fig.~\ref{fig:FM_band}, we present the magnon dispersion with and without the phonon-induced IL-DMI. In general, the phonon-induced IL-DMI favors a co-planar canted magnetic groundstate. Depending on the strength of the interlayer Heisenberg exchange, the middle two magnon bands may cross. However neither an external magnetic field, Fig.~\ref{fig:FM_band}(b), nor an IL-DMI along can cause an avoided crossing. Only their combination induces non-coplanar canting and opens a gap between the middle bands, as seen in Fig.~\ref{fig:FM_band}(d), and also Fig.~\ref{fig:AFM_band} in Appendix~\ref{more_dispersion}.

\subsection{Magnon topology and band Chern number}

With this transformation the problem of solving the BdG Hamiltonian has been reduced to solving the eigenproblem of the non-Hermitian Hamiltonian $\Sigma_z H_{BdG}$. The positive eigenvalues are the physical magnon spectrum. The non-Hermiticity modifies the inner product in this symplectic vector space $ \langle \cdot \; , \; \cdot \rangle: V \times V  \rightarrow \mathbb{R}$ as $\langle \phi,\psi\rangle:=\phi^\dagger\Sigma_z\psi$ which in turn modifies the definition of the magnon Berry connection and Berry curvature \cite{Kondo2020Non-HermiticitySystems},
	\begin{equation}
	    \vec A_{n\tau}(\vec{k})=i\tau\nu^\dagger_{n\tau}(\vec{k})\Sigma_z\vec \nabla_{\vec k}\nu_{n\tau}(\vec{k}),
	\end{equation}
where $\nu_{n\tau}(\vec k)$ is the eigenvector of $\Sigma_z H_{BdG}(\vec k)$ with eigen-energy $E_{n,\vec k}$, $\tau=+(-)$ for particle(hole) bands. The Berry curvature is given by
	\begin{equation}
	    \vec \Omega_{n\tau}(\vec{k})=\nabla_{\vec{k}}\times \vec A_{n\tau}(\vec{k}).\end{equation}
  The numerical evaluation of the Berry curvature is done following Fukui's method \cite{Fukui2005ChernConductances}.  The integration of the Berry curvature over the first BZ is the Chern number,
\begin{equation}
\mathcal{C}_{n\tau}=\int_{1^{st}BZ}  \Omega_{n\tau}(\vec k)d\vec k^2,
\end{equation}
a topological invariant of the magnon bands which can only change with gap closings and re-openings.

Because a gap opens with non-zero IL-DMI in the presence of a magnetic field, it is natural to investigate the topological properties through magnon band Chern numbers. The IL-DMI gives an extra tuning knob of the system's properties that can be controlled with ultrafast precision using a laser drive. Some choices of Hamiltonian parameters with a gapped magnon spectrum and non-zero Chern numbers are listed in Table \ref{table-magnon}. 

Note that when $\vec B=0$, the interlayer FM system has two degenerated groundstates, one with all spins up and one with all spins down. These two have opposite Chern numbers, so we assume a small infinitesimal symmetry breaking field to pick the all spins up groundstate. A similar situation happens in interlayer AFM coupling case. Although both cases share the same Chern number, we still assume a symmetry-breaking field to pick one of the groundstates: canted from up for the top layer, and canted from down for the bottom layer. In the FM case, when a finite external field is applied, we allow the ground state to automatically follow the minimal energy configuration. The coexistence of an IL-DMI and an external B field is essential to gap out the middle two bands in the AFM and FM cases and enable a correct assignment of Chern number to each band. 

Temperature-dependent thermal Hall measurements are one possible route to experimentally gain indirect access to the Chern numbers and Berry curvatures of topological magnon bands \cite{PhysRevB.98.094419,PhysRevLett.118.177201,PhysRevB.105.L100402}.  Because the population of the magnon bands depends on the temperature, the lowest-lying bands will determine the thermal transport properties \cite{PhysRevLett.118.177201}. However, since magnons are bosons, it is not possible to have completely ``filled bands," and therefore, a quantized thermal Hall response is not expected (as it would be for fermions that completely occupy some set of bands). While transport properties are not the focus of this paper, we note here that it is possible to partially infer information about the Berry curvature and Chern numbers from the thermal Hall response. While such ``fast" (on the scale of nano or even picoseconds) transport experiments are extremely challenging, recent progress in electron transport shows that capabilities are rapidly advancing \cite{McIver:np2020}.  Finally, we remark that magnetically sensitive light experiments (e.g., through frequency-dependent light polarization rotation) could also be used to infer some properties of the magnon band structure. 

\begin{table}[h]
\begin{center}
\begin{tabular}{ | m{3cm} || m{1.3cm}| m{2.1cm}|m{1.5cm} | } 
  \hline
  Hamiltonian Parameters& FM $J_2=-0.2$  &FM $J_2=-0.3$  & AFM $J_2=0.3$  \\ 
  \hline \hline
  $\vec D_i=0$, $\vec B=(0,0,0)$ &-1,-1,1,1 &-1,ND,ND,1  & degenerate \\ 
  \hline
  $\vec D_i=0$,  $\vec B=(0.2,0.2,0.2)$& -1,-1,1,1,& -1,ND,ND,1 & -1, 1, 1, -1\\ 
  \hline
  $\vec D_i=-0.2\hat{x}$, $\vec B=0$&-1,-1,1,1, & -1, ND,ND,1  & degenerate\\
  \hline
  $\vec D_i=-0.2\hat{x}$, $\vec B=(0.2,0.2,0.2)$&-1,-1,1,1, & -1, 1,-1,1  & -1, 1, 1, -1\\
   \hline
  $\vec D_i=-0.2\hat{x}$,  $\vec B=(-0.2,-0.2,-0.2)$& 1,1,-1,-1,& 1,-1,1,-1 & 1, -1, -1, 1\\ 
  \hline
\end{tabular}
\caption{Table of magnon particle band Chern numbers for various system parameters. The Chern numbers (either +1 or -1) for the magnon particle bands are expressed as $\mathcal{C}=\{\mathcal{C}_{1,+},\mathcal{C}_{2,+},\mathcal{C}_{3,+},\mathcal{C}_{4,+} \}$. For all the calculations shown in this table, we have used $J_1$=-1, $\Delta=-0.8$, and $D_z=0.05$. ND means that the Chern numbers are not well defined when there are band touching points between bands. In the case of $\vec B=\vec 0$, a small infinitesimal field is applied to choose the all spin up groundstate in the FM interlayer coupling case.} 
\label{table-magnon}
\end{center}
\end{table}

\subsection{Relevant materials platforms}
Our focus is primarily on the possibility of an ultrafast on-off switch of the IL-DMI and thus we mainly focused on the AA stacking and AB stacking whose IL-DMI are zero in the equilibrium settings. However, we note that bilayer honeycomb lattices can have AB' stacking order\cite{sivadas2018stacking, handy1952structural}, and this is relevant to bilayer CrI$_3$, see Fig.(\ref{fig:AB'}) which is easier to achieved in experiment then the AA stacking of bilayer CrI$_3$. Such AB' stacking order of bilayer CrI$_3$ actually has a non-zero IL-DMI in the static configuration, yet the nonlinear phononics can still modify the strength of the IL-DMI. We have discussed this situation in more detail in Appendix~\ref{third_stacking},   although the exact value of the IL-DMI strength would need to be evaluated from first-principle methods\cite{stavric2023delving, yang2023first}. 

In general, there are many bilayer van der Waals magnets that are interesting material platforms for using nonlinear phononics to tune the IL-DMI. To list a few, bilayer CrBr$_3$ and CrCl$_3$, in the same materials family\cite{kim2019evolution},  have slightly different magnetic couplings that might prove interesting to see how IL-DMI could give rise to different magnetic orders, including under a light drive. Also, given the recent interest in twistronics\cite{bistritzer2011moire}, some twisted bilayer systems such as twisted bilayer CrI$_3$\cite{song2021direct}, twisted bilayer graphene\cite{lu2019superconductors} and twisted hetero\cite{li2021quantum}- or homo\cite{cai2023signatures}-bilayer transition metal dichalcogenides (TMDs), which host interesting magnetic orders (although they sometimes compete or intertwine with electronic orders, such as integer and fractional quantum anomalous Hall\cite{park2023observation}), may be interesting to see how IL-DMI can change the magnetic orders. While the details of these materials are beyond the description of our toy models, the  same methods can be applied, in combination with first principles calculations \cite{stavric2023delving, yang2023first}, to study them.

\section{Conclusion and Outlook}
\label{sec:conclusion}

In this work, we have theoretically studied a bilayer honeycomb lattice model subjected to a strong laser drive in resonance with infrared phonons. In this regime, phonon anharmonic couplings are relevant, and transient lattice distortions can be induced via cubic infrared-Raman phonon couplings. We further showed that the transiently distorted lattice possesses lower symmetry than an equilibrium lattice and that an interlayer Dyzaloshinkii-Moriya interaction is allowed in an AA stacked system. This is one of the central results of this work. 

We then explored the resulting changes in the magnetic Hamiltonian, magnetic ground states, and excitations above the ground state. We found that the interlayer Dzyaloshinkii-Moriya interaction produces spin canting in the ground state which can endow the magnetic excitations (magnons) with topological properties. We explicitly computed the Berry curvature and Chern numbers of the magnon bands for selected parameters and demonstrated that topological transitions occur, such as when a static external magnetic field is applied.

For technological applications, it is beneficial to have control and tunability of material properties.  We have discussed the advantages of using phonon-photon interactions to engineer magnetic systems (ultra-fast reversible control of the magnetic Hamiltonian). For example, because the energy scale of the phonon-photon interaction is smaller than the electron-photon one, and because the phonon energy scale is typically lower than the electronic energy scale, directly exciting phonons produces fewer undesirable heating effects in the system. 

Our work serves as a proof of concept for the nonlinear phononics mechanism for modifying interesting magnetic phases and their excitations through the concrete protocol for inducing an interlayer Dyzaloshinkii-Moriya interaction in layered van der Waals materials.  We hope this work will inspire further theoretical and experimental efforts in this direction.

\begin{acknowledgments}
Z.L. is grateful to Michael Vogl and Luyan Yu for their helpful discussions and inspirational comments. Special thanks to Benjamin Wieder for his detailed guidance and discussions on symmetry, topology and everything condensed matter physics. This research was primarily supported by the National Science Foundation through the Center for Dynamics and Control of Materials: an NSF MR- SEC under Cooperative Agreement No. DMR-1720595 and NSF Grant No. DMR-2114825.  G.A.F acknowledges additional support from the Alexander von Humboldt Foundation.
\end{acknowledgments}

\newpage

\bibliography{magnonproject}

\newpage

\appendix
\newpage

\section{Group Theory Analysis of Phonons}\label{append:GroupTheory}
	The lattice vibration modes can be determined by group theory from \cite{Dresselhaus2010GroupMatter},
	\begin{equation}
	  \chi_{vibration}=\Gamma^{equivalence}\otimes\Gamma_{vec},  
	\end{equation}
where $\Gamma^{equivalence}$ counts the number of atoms within one unit cell which are mapped to themselves under spatial operations, and $\Gamma_{vec}$ is the sum of the modes that transform as vectors.

The AB stacked honeycomb has a $D_{3d}$ point group,
\begin{equation}
	  \chi_{vibration}=2A_{1g}\oplus2A_{2u}\oplus2E_g\oplus2E_u,
	\end{equation}
and the AA stacked honeycomb has a $D_{6h}$ point group,
\begin{equation}
	  \chi_{vibration}=A_{1g}\oplus A_{2u}\oplus B_{2g}\oplus B_{1u}\oplus E_{2g}\oplus E_{1u}\oplus E_{1g}\oplus E_{2u}.
	\end{equation}
With this representation, one can determine the vibrational motion in real space \cite{Rodriguez-Vega2020Phonon-mediatedI3,Rodriguez-Vega2022Light-DrivenAntiferromagnets}.  Both AA and AB stacked honeycomb lattices have twelve phonon modes since there are four atoms in the unit cell.

 \begin{figure}[ht!]
\includegraphics[width=50mm]{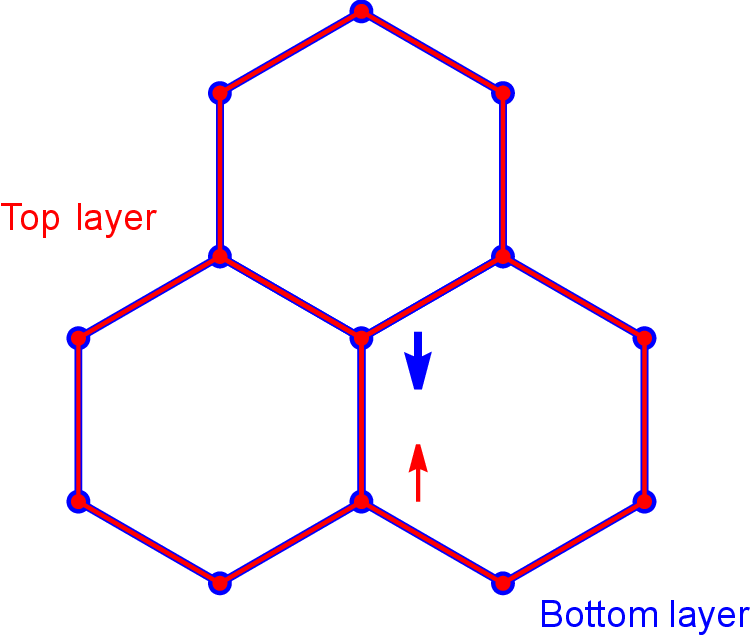}
\caption{\label{fig:E1g} $E_{1g}$ mode of AA stacked bilayer honeycomb lattice in which atomic sites sit directly above or below one another.  Here the $E_{1g}$ mode exhibits oscillations producing a shearing motion in the plane of the bilayer as indicated by the arrows, where the blue arrow means the bottom layer shears in the ``downwards" direction and the red arrow means the top layer shears in the ``upwards" direction. All the atoms in one layer move together in the same phase.}
\end{figure}

\begin{figure}[ht!]
\includegraphics[width=50mm]{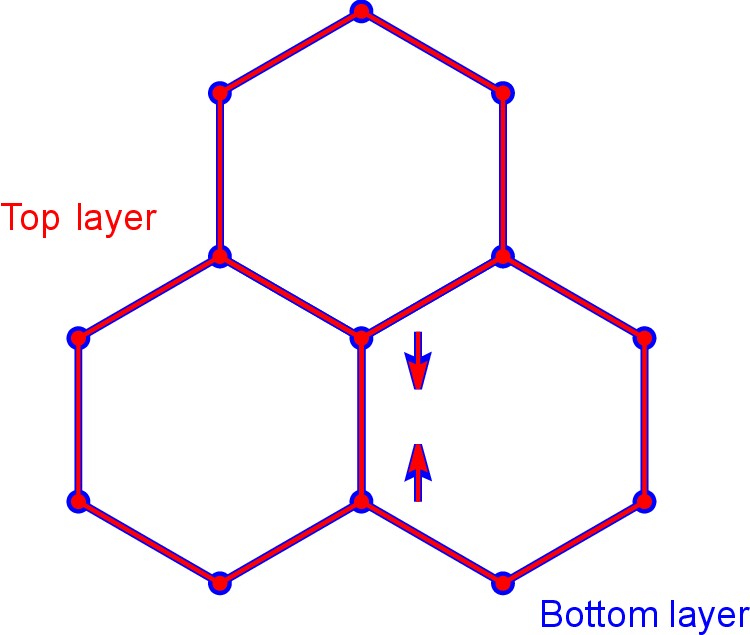}
\caption{\label{fig:E2g2} $E_{2g}$ oscillation in the $y$-direction for an AA stacked honeycomb bilayer, effectively changing the bonds length in $y$ direction, where the B sublattice of both layers moving towards $-y$ and A sublattice of both layers moving towards $y$.}
\end{figure}

The $E_{1g}$ modes describe shearing between layers \cite{Moriya1960AnisotropicFerromagnetism}. The shearing modes break the $xy$-plane mirror symmetry and all two-fold symmetries, as well as the three-fold symmetry in the $z$-direction normal to the plane of the bilayer. Therefore, a nearest neighbour DMI between the two closest interlayer atoms, one from the top layer and one from the bottom, is no longer forbidden.  Symmetry requires that the DM vector be perpendicular to the $yz$ plane.

\begin{figure}[ht]
\includegraphics[width=50mm]{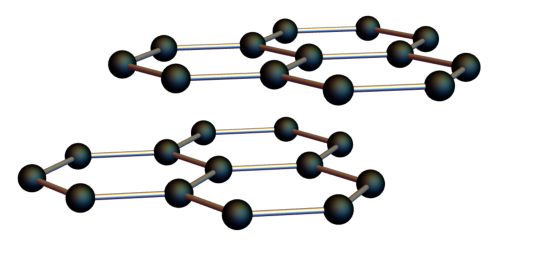}
\caption{ AB stacking of a bilayer honeycomb lattice.  The top layer (shown in red) lies above a bottom layer (shown in blue) that is shifted so that the atoms of the top layer sit above the centers of the hexagons in the bottom layer.}
\label{fig:abstacking}
\end{figure}

To see what kind of phonon interaction is allowed, we need to see if such interaction respect to crystal symmetry. For example in our case, we have $A_{1g}\subseteq E_{1u}^2\otimes E_{2g}$, meaning that $Q_{IR}^2 Q_{R}$ is an allowed interaction between phonons. More specifically, we have
\begin{equation}
   V_{anharmonic}=({Q_{IR}^y}^2-{Q_{IR}^x}^2)Q_{R}^a+2Q^x_{IR}Q^y_{IR}Q_{R}^b,
\end{equation}
where $Q_{IR}$ is a IR mode with $E_{1u}$ irrep and  $Q_{R}$ is a Raman mode with $E_{2g}$ irrep.

\subsection{Real space displacement of phonon}
The projection operator is constructed in the following way: \cite{Dresselhaus2010GroupMatter}
 \begin{equation}
    \hat P^{(\Gamma_n)}_{kl} = \frac{l_n}{h} \sum_{C_\alpha} \left( D_{kl}^{(\Gamma_n)}(C_\alpha) \right)^* \hat P(C_\alpha), 
 \end{equation}
where $D_{kl}^{(\Gamma_n)}(C_\alpha)$ is the irreducible matrix
representation of the group element $C_\alpha$, $h$ is the order of the group, $l_n$ is the dimension of the irreducible representation, and $\hat P(C_\alpha)$ is the representation of $C_\alpha$ constructed by the permutation matrix and the O$(3)$ symmetry operations.
Once, we have this, the eigenmode for irreps $\gamma_n$ can be calculated by finding the eigenvector of operator $\hat P^{(\Gamma_n)}_{ll}$. 
In $E_{2g}$ mode, the lattice has the following displacement vectors: 

\begin{equation}
    E_{2g,x}: (\Delta Q^t_a, \Delta Q^t_b,\Delta Q^b_a, \Delta Q^b_b)=(-\vec i,\vec i, -\vec i,\vec i)
\end{equation}
\begin{equation}
    E_{2g,y}: (\Delta Q^t_a, \Delta Q^t_b,\Delta Q^b_a, \Delta Q^b_b)=(-\vec j,\vec j, -\vec j,\vec j)
\end{equation}
where $\Delta Q^{t(b)}_a$, $\Delta Q^{t(b)}_b$ are the displacement vectors for $a,b$ sublattices in t(b) layer in the unit cell respectively, $\vec i$ and $\vec j$ are unit vectors in the $x$ and $y$-directions.

\subsection{Pictorial understanding of the 3-fold symmetry breaking}

\begin{figure}[h!]
		\includegraphics[width=60mm]{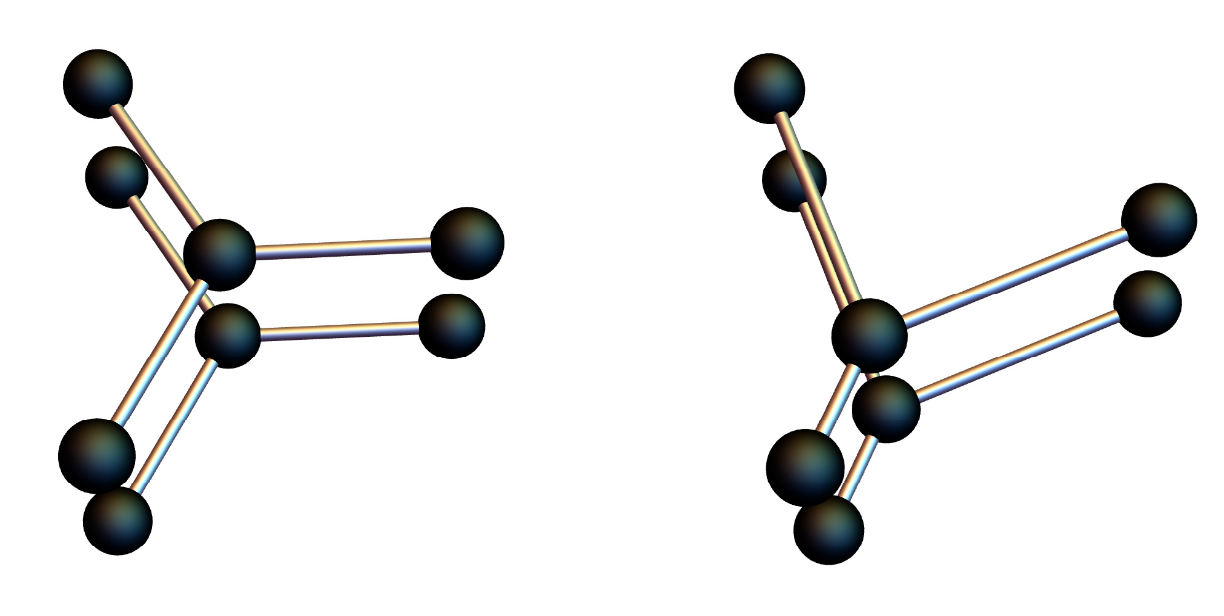}
		\caption{(Left): Interlayer DM vector $\vec D=0$ when $Q_R=0$. (Right): Interlayer DM vector is nonzero when $Q_R\neq0$, due to the breaking of $C_3$ rotational symmetry}
	\label{fig:DM_QR_0}
\end{figure}

A 3-fold rotational symmetry guarantees a cancellation of all the contributions to the IL-DMI.

\section{Symmetry analysis on IL-DMI}\label{sym_DMI}

\subsection{Comments on magnetic exchange  parameters}
In the main text, we chose the nearest neighbor bonds to have the same exchange interaction $J_{1a}=J_{1b}$. Although in the driven case this is only an approximation, the key physics remains the same.

\subsection{Some comments on interlayer DM interaction}

\subsubsection{Moriya's rules}
Here we list the original Moriya's rules on this antisymmetric spin coupling \cite{Moriya1960AnisotropicFerromagnetism}. Following the same notation used by Moriya, the coupling between two ions in the crystal is considered, and these two ions 1 and 2 sit at position A and B, the midpoint of AB is denoted as C. Moriya showed the following rules apply:
\begin{enumerate}
    \item When a center of inversion is located at C, then $\bm D=0$.
    \item When a mirror plane perpendicular to AB passes through C, then $\bm D \parallel$ mirror plane or $\bm D \perp$ AB.
    \item When there is a mirror plane including A and B, then $\bm D \perp$ mirror plane. 
    \item When a two-fold rotation axis  perpendicular to AB passes through C, then $\bm D \perp$ two-fold axis.
    \item When there is an $n$-fold rotation axis ($n\geq 2$) along AB, then $\bm D \parallel$ AB.  
\end{enumerate}

\subsubsection{Application of Moriya's rules to undistorted lattice}

We now apply these rules to our AA stacked bilayer honeycomb. ($D_{6h}$). In this case, when one looks at the interlayer coupling between ion 1 (denoted position A) on the top layer and ion 2 (denoted position B) directly below ion 1, Moriya's rules apply as follows. According to Rule 5, there is a 3-fold rotation axis along AB, thus $\bm D \parallel$ AB and $\perp$ to the plane of the material. 
 According to Rule 2, there is a mirror plane perpendicular to AB passing through C, thus $\bm D \parallel$ mirror plane or $\bm D \perp$ AB, which contradicts with the conclusion of Rule 5 unless $\bm D= \bm 0$. One also sees that Rule 3 and 4 would rule out the possibility of a $\bm D \parallel$ AB.

\subsubsection{Application of Moriya's rules to the non-equilibrium distorted lattice}

When the $E_{2g,x(y)}$ phonon mode is excited, the 3-fold rotation axis along AB is broken, so Rule 5 does not apply.  Since there is no center of inversion in C, we now only have the restrictions of Rules  2, 3, and 4. We now consider the $E_{2g, y}$ stretching along the bond, and find $\bf D$ $\perp$ AB and $\perp$ to the $yz$-plane.

\begin{figure}[h!]
		\includegraphics[width=30mm]{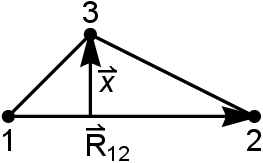}
		\caption{DM interaction model relevant to the atomic position shifts associated with an $E_{2g}$ mode oscillating along the $y$-direction. }
		\label{fig:DM_fig}
\end{figure}

A simple model \cite{Cheong2007Multiferroics:Ferroelectricity} that takes into account a third atom that breaks the inversion symmetry is illustrated in Fig.\ref{fig:DM_fig}, where a $DM$ vector	
\begin{equation}
    \vec D_{12}\propto \lambda \vec{x}\times \vec{r}_{12},
\end{equation}
is generated from inversion symmetry breaking where $\lambda$ is a coefficient reflecting the strength of spin-orbit coupling, $\vec r_{12}$ is a vector connecting atom 1 and atom 2, $\vec x$ is the perpendicular to $\vec r_{12}$ that points toward the third atom. In our situation, atom 1 and 2 are located in the opposite layer, and atom 3 would be some environmental atom.

\begin{figure}[h!]
		\includegraphics[width=0.8\linewidth]{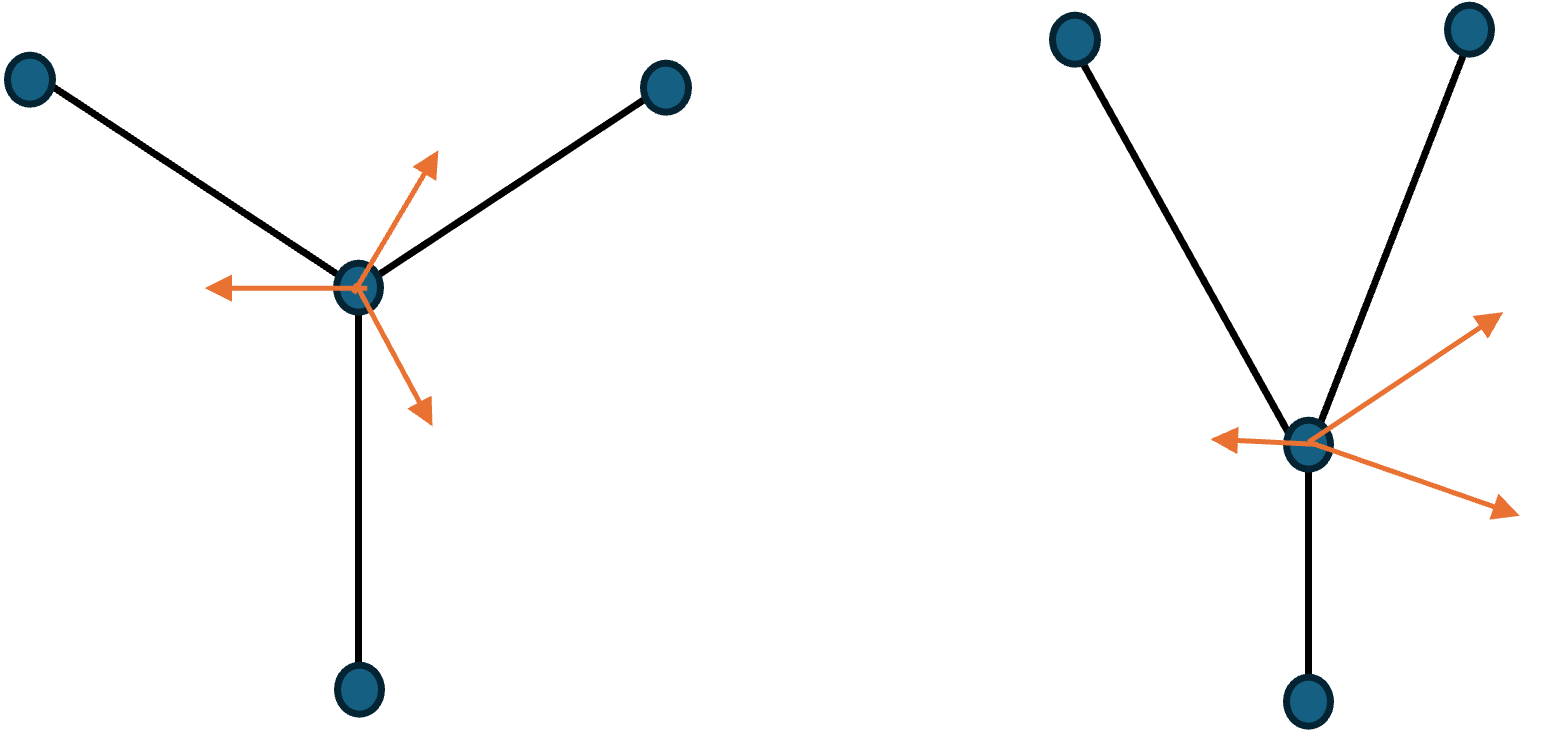}
		\caption{IL-DMI has contributions from three NN atoms. (left) DM vector $\vec D_{12}=0$ when $Q_R=0$, the arrows show the IL-DMI between the top and bottom atoms from the mediations of neighboring atoms.  (right) DM vectors when $Q_R\neq0$, the arrows show the DM vector contribution from neighboring atoms. }
	\label{fig:IL_DMI}
	\end{figure}

\subsection{Comments on AB' stacking order}
\label{third_stacking}

\begin{figure}[htp]
    \centering
    \includegraphics[width=1\linewidth]{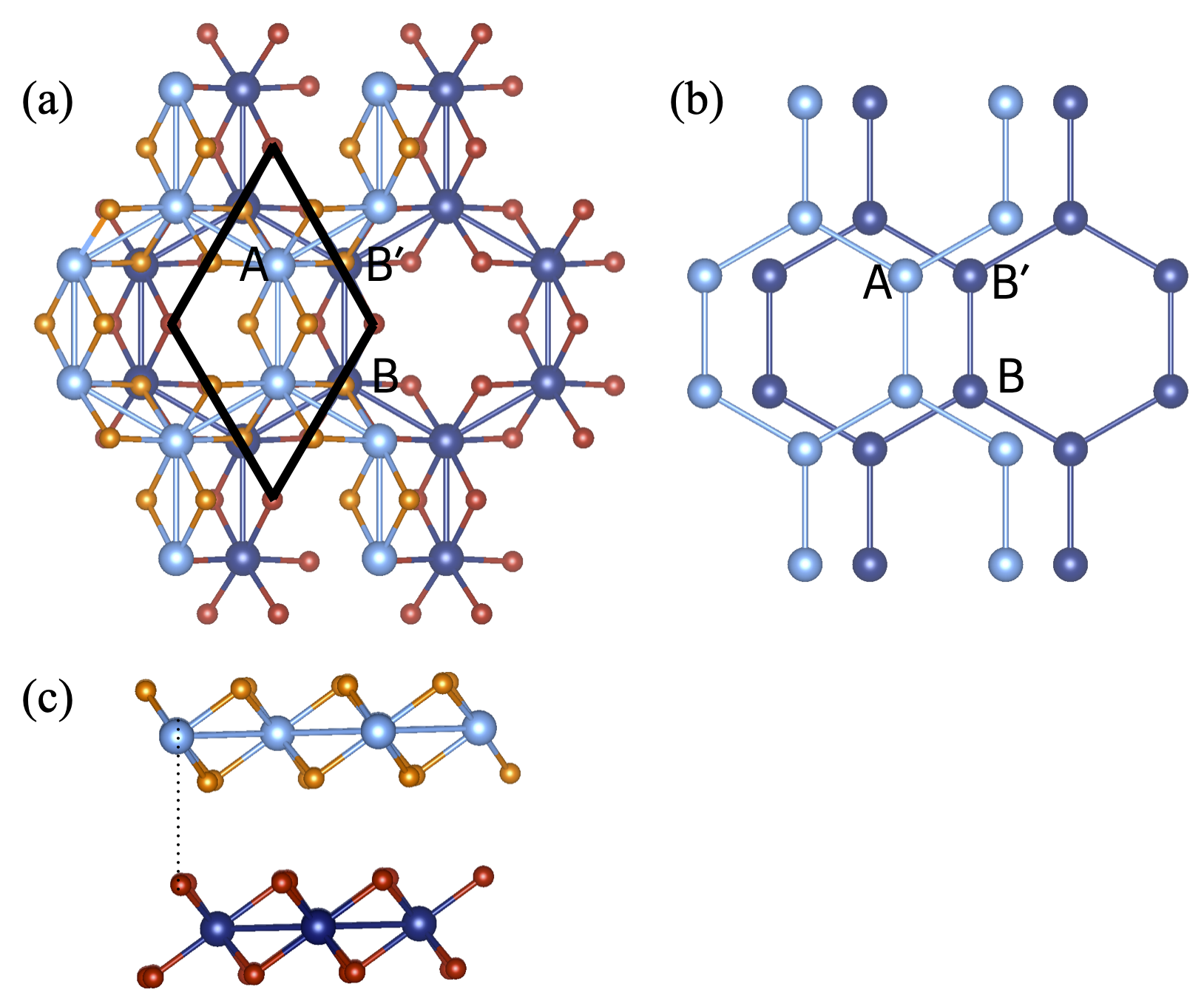}
    \caption{Crystal structure of AB' stacking bilayer $CrI_3$, $C2/m$ space group, $C_{2h}$ point group. (Plotted with VESTA\cite{momma2011vesta}) (a) Top view. (b)Top view with magnetic ions only. (c) Side view.}
    \label{fig:AB'}
\end{figure}

The main text of our paper focuses on the possibility of an ultrafast on-off switch of the IL-DMI and thus mainly focuses on the AA stacking and AB stacking whose IL-DMI are zero in the equilibrium settings. However, a bilayer honeycomb lattice can have AB' stacking order\cite{sivadas2018stacking, handy1952structural}, and this is relevant to bilayer CrI$_3$, see Fig.\ref{fig:AB'} which is easier achieve in experiment. In this point group symmetry, one can apply Moriya's rules\cite{Moriya1960AnisotropicFerromagnetism} and comment on its IL-DMI. 

AB' stacking corresponds to a fractional lateral shift from AB stacking. In the AB stacking, we denote the top ion as A and the bottom ion as B. There is an inversion center in the midpoint of AB, which we denote as C. After the lateral shift, the A ion is closer to another ion in the bottom layer, denoted as B'. According Moriya's rules, there can be no interlayer DMI between A and B. However, an interlayer DMI is permitted between A and B'. The reason is the following: there is no center of inversion between A and B'. There is no mirror plane $\perp$ AB', and no mirror plane including AB'. There is also no n-fold rotation axis along AB'. It only has a 2-fold rotation axis $\perp$ AB' passing through C, so this is the only restriction on $\bm D$, which states  $\bm D \perp$ to such a two-fold rotation axis. Therefore, we know that by symmetry constraint alone, IL-DMI is generally allowed.  In this case, while it is non-zero, the nonlinear phononics protocal could control the strength of the IL-DMI. However, to what degree the nonlinear phononics control this IL-DMI becomes a question of the energy scales, such that a detailed answer requires a first-principle calculation\cite{stavric2023delving, yang2023first} and is beyond the scope of our current discussion.

\section{Berry curvature and energy dispersion of magnon for AFM interlayer coupling}
\label{more_dispersion}
\begin{figure}[htp]
    \centering
\includegraphics[width=0.48\textwidth]{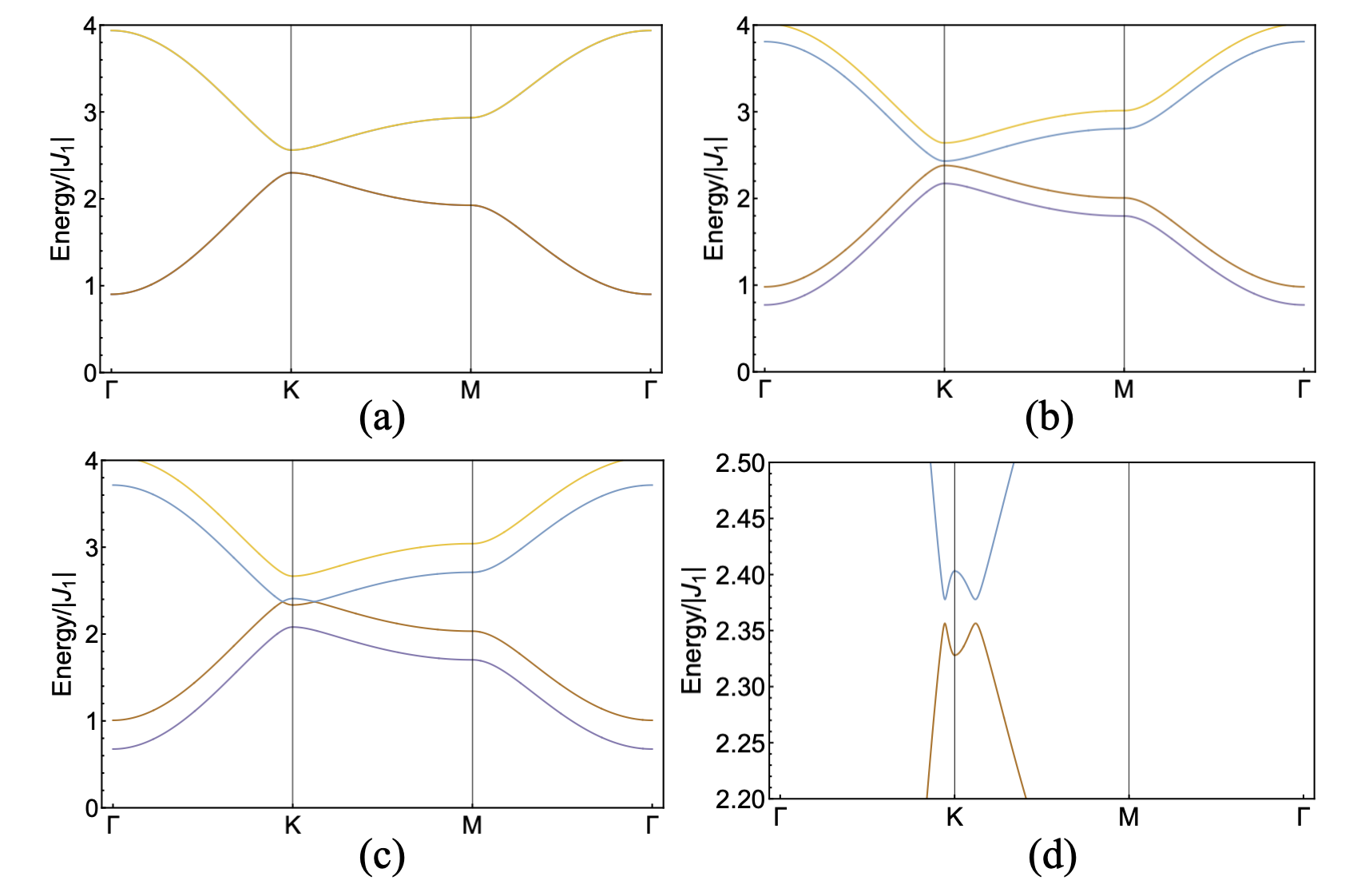}
    \caption{Magnon spectrum with AFM interlayer coupling $J_2=0.3$, other model parameters are: intralayer exchange couplings $J_1=-1$, easy axis anisotropy $\Delta=-0.8$, intralayer DMI strength $D_z=0.05$, (a) IL-DMI  $\vec D_i=-0.1$, $\vec B=(0,0,0)$, (b)IL-DMI $\vec D_i=0 \hat x$, $\vec B=(-0.2, -0.2, -0.2)$, (c)gap clossing between middle two bands with increasing magnetic field, plotted with $\vec D_i=0 \hat x$, $\vec B=(-0.3, -0.3, -0.3)$. Here, (a) and (b)(c) show the band structures of the cases with only the phonon-driven IL-DMI applied or an external magnetic field applied, respectively. (d) show a gap opening between the two middle bands when we have a combination of both the IL-DMI and an external magnetic field. }
        \label{fig:AFM_band}
\end{figure}

\begin{figure}[h!]
		\includegraphics[width=85mm]{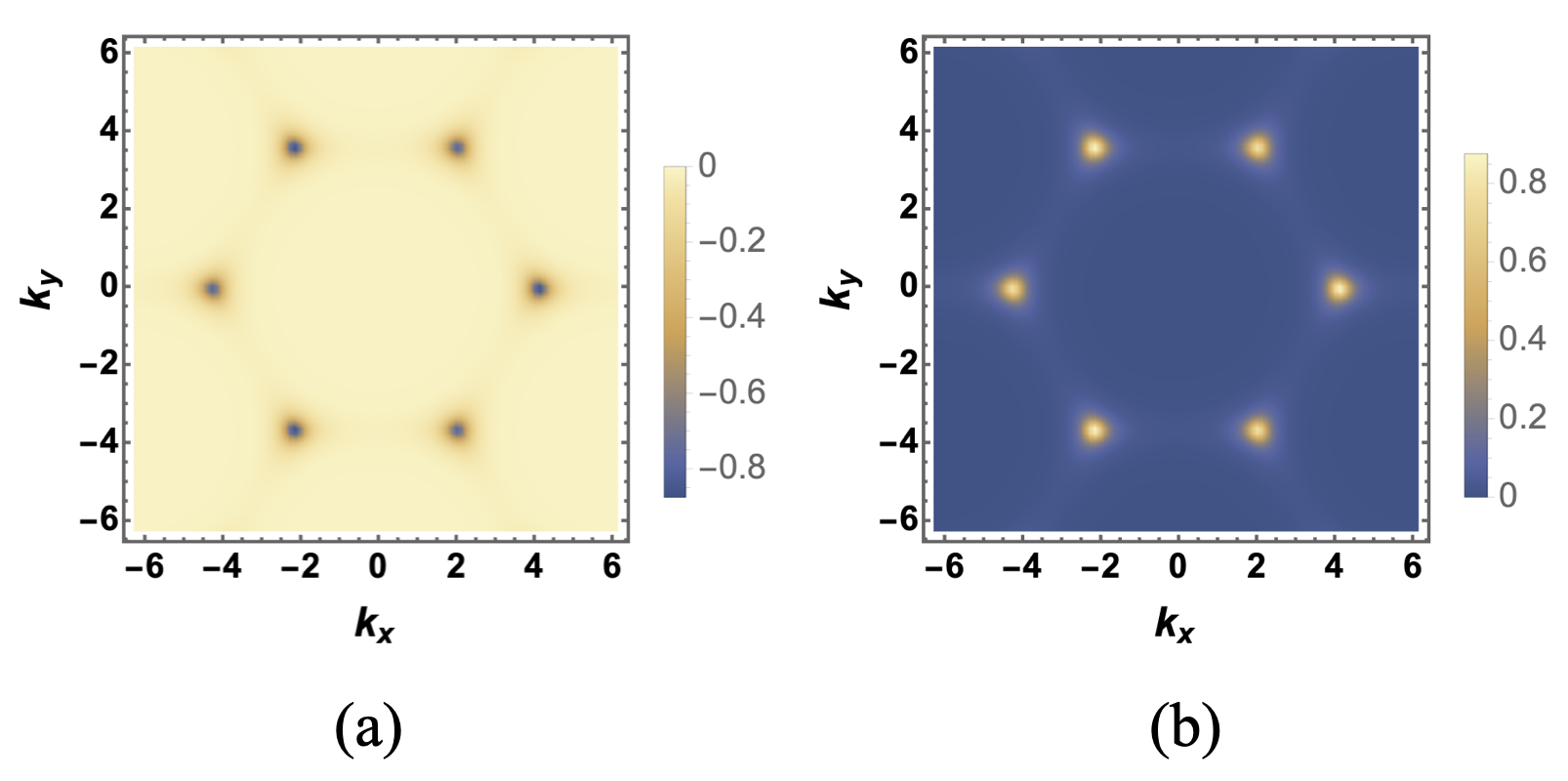}
		 \caption{Berry curvature for the third magnon band with AFM interlayer coupling $J_2=0.3$, other model parameters are: intralayer exchange couplings $J_1=-1$, easy axis anisotropy $\Delta=-0.8$, intralayer DMI strength $D_z=0.05$, . The combination of IL-DMI and external magnetic field produces a gap between the middle two bands and ensures well-defined Chern numbers. External magnetic field switches the signs for the Chern number of each band. (a) shows the berry curvature distribution of a Chern -1 band, with IL-DMI $\vec D_i=-0.1 \hat x$, $\vec B=(-0.2, -0.2, -0.2)$, (b) shows the berry curvature distribution of a Chern 1 band, with IL-DMI $\vec D_i=-0.1 \hat x$, $\vec B=(0.2, 0.2, 0.2)$. }
        \label{fig:berry_group}
\end{figure}

Berry curvature and energy dispersion of magnon are reported in Fig.~\ref{fig:AFM_band} and Fig.~\ref{fig:berry_group}.

\end{document}